# Empirical Study of DSRC Performance Based on Safety Pilot Model Deployment Data

Xianan Huang, Ding Zhao, Huei Peng

*Abstract*— Dedicated Short Range Communication (DSRC) was designed to provide reliable wireless communication for intelligent transportation system applications. Sharing information among cars and between cars and the infrastructure, pedestrians, or "the cloud" has great potential to improve safety, mobility and fuel economy. DSRC is being considered by the US Department of Transportation to be required for ground vehicles. In the past, their performance has been assessed thoroughly in the labs and limited field testing, but not on a large fleet. In this paper, we present the analysis of DSRC performance using data from the world's largest connected vehicle test program—Safety Pilot Model Deployment lead by the University of Michigan. We first investigate their maximum and effective range, and then study the effect of environmental factors, such as trees/foliage, weather, buildings, vehicle travel direction, and road elevation. The results can be used to guide future DSRC equipment placement and installation, and can be used to develop DSRC communication models for numerical simulations.

## I. Introduction

Dedicated short-range communication (DSRC) supports short range and reliable data communication between vehicles and with infrastructures to enable a number of safety, mobility, and energy applications [1]. A set of Basic Safety Messages (BSM) has been defined in SAE J2735 to ensure safety critical messages such as vehicle position, velocity, headway, and deceleration are broadcasted at a frequency of 10Hz. In the European Union, the Cooperative Awareness Messages (CAMs) have been specified in an ITS standard [2].

Extended from IEEE 802.11a wireless communication protocol, which was designed for short range, low mobility, indoor use, IEEE 802.11p[3] is designed to meet the requirements of longer range (up to 1 km), extremely high mobility, and rapidly changing channel conditions. Because of these differences, DSRC operates at a higher frequency band, between 5.850 GHz and 5.925 GHz. This band is "dedicated" by the US Federal Communications Commission (FCC) for intelligent transportation system applications, to reduce the possibility of interference with other wireless devices [4]. Since one of the major purposes of DSRC is safety, reliable communication is critical.

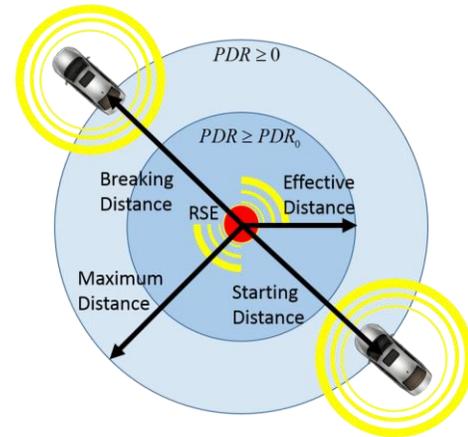

Fig. 1. Maximum range and effective range of DSRC communication

To characterize the performance of DSRC, the following metrics [5] are defined:

1) Packet Delivery Ratio (PDR): the ratio of successful communication events to the total number of transmission attempts at a given distance between two DSRC units.

2) Maximum Range (MR): the maximum distance at which the vehicle or road side equipment (RSE) can receive packets from another vehicle with a larger-than-zero packet delivery ratio.

3) Effective Range (ER): the distance within which the vehicle or RSE can receive packets from other vehicles with a packet delivery ratio larger than a defined threshold (e.g., 50%).

The metrics are shown in Fig. 1. At the center of the circle is an RSE, which is frequently installed at strategically selected locations such as at a road intersection. For any vehicle equipped with a DSRC, it continues broadcasting BSM which may successfully reach the RSE. The starting distance is defined as the distance where the RSE receives the first message from the vehicle, and the breaking distance is defined as the distance where the RSE receives the last message from the vehicle. These two values from each trip are used to calculate the RSE's maximum range and effective range when the vehicles move towards or away from the RSE.

The contribution of this paper includes:
1) We presented the DSRC performance under real-life driving condition from a large scale naturalistic driving database, which makes the evaluation close to the ground truth of DSRC performance in future deployment.
2) We evaluated the performance of DSRC under the influence of a series of dominant real-life factors including weather,

foliage, road elevation, aftermarket supplier, etc. using data under influence if multiple lumped factors.

The main objective of this paper is to evaluate the performance of DSRC, including maximum range and packet delivery ratio. The rest of the paper is organized as follows. Section II summarizes the related works. Section III describes the naturalistic driving database used in the analysis. Section IV shows the analysis results. Section V then concludes the paper.

## II. RELATED WORKS

Our work is inspired by a number of prior studies on the performance of DSRC. When IEEE 802.11p was published in 2010 and the DSRC characteristics were unclear to the Vehicular Ad-hoc Network (VANET) community, Bai et al. [6] studied the influence of controlled factors such as radio parameters and uncontrolled factors such as distance and vehicle velocity on the DSRC packet delivery ratio. The behavior of DSRC under vehicular blockage was studied statically in a parking lot as well as dynamically in urban, suburban and highway scenarios [7]. A recent work [5] also used logged data with multiple testing vehicles to study the performance of DSRC as a function of the distance between the transmitter and the receiver. Controlled lab testing is also useful in studying the congestion scenario [8] because a high number of transmitters can be easily achieved under lab setting and other variables can be controlled properly.

Simulations are sometimes used to study the influence of traffic density to DSRC due to the cost of deploying VANET on a large scale. Schumacher et al. [9] compared different models for DSRC propagation such as log-distance, two-ray ground reflection and dual slope models, and developed a 1-D highway propagation model mainly considering the Nakagami model with empirical signal strength. Ma et al. [10] studied the influence of packet collision and hidden terminal problems with a 1-D one-hop broadcast VANET model and showed the packet delivery ratio and effective range under the influence of traffic density, packet generation rate and data transmission rate. Although used widely in DSRC related research, the accuracy of channel model under real-world condition remains questionable. In order to validate the DSRC model, Biddlestone et al. [11] derived a propagation model for urban environment and validated the model with experimental data from a testing track. Analytical methods have also been used to study the influence of the medium access control (MAC) protocol and radio parameters [12]-[14].

Unlike controlled experiments, naturalistic driving contains unknown and uncontrolled real-world driving conditions such as weather, traffic densities, road geometries, etc. Recently, naturalistic driving data becomes available and was used in the evaluation of vehicle safety systems such as roll stability control and lane departure warning, as well as for the analysis of driver behaviors [15]-[17]. Naturalistic driving databases contain a large volume of diverse results which better reflect real-world situations and challenges. Some factors such as the influence of road elevation changes, trees and weathers are better reflected in these databases than in a controlled lab testing. Most importantly, the results better reflect real-world performance of DSRC.

## III. EXPERIMENT DATA DESCRIPTION

### A. Naturalistic Driving Database

The data used is from the Safety Pilot Model Deployment (SPMD) project lead by the University of Michigan Transportation Research Institute (UMTRI). SPMD has as many as 2,800 passenger cars, trucks and buses equipped with devices for V2V and V2I communication. For the infrastructure side, there are 25 roadside equipment (RSE), 21 at signalized intersections, the remainder at curves and freeway locations. The experiment has been running continuously since August 2012 for more than 1,000 days, and has collected more than 5.6 TB of recorded Basic Safety Messages (BSM) [18].

There are four different types of vehicle equipment configurations in SPMD vehicles, referred as Integrated Safety Device (ISD), Aftermarket Safety Device (ASD), Retrofit Safety Device (RSD), and Vehicle Awareness Device (VAD). The configurations are summarized in TABLE I.

TABLE I
SPMD DSRC Device Summary

| Device | Tx | Rx | Weight Class | Quantity | Supplier |
|---|---|---|---|---|---|
| ISD | Y | Y | Light | 67 | A |
| VAD | Y | N | Light, Medium, Heavy Duty, Transit | 2450 | B, C |
| ASD | Y | Y | Light | 300 | A, B |
| RSD | Y | Y | Heavy Duty, Transit | 19 | A, B |

Among the 300 vehicles equipped with ASD, 98 are equipped with data acquisition system (DAS) custom made by UMTRI, which is used to record data such as forward object information, position information, lane tracking information, and remote vehicle BSM and classification. Because of the requirement that no permanent modification should be made to the vehicle, for most of the vehicles equipped with VAD or ASD, the DSRC antenna is installed on the rear cargo shelf inside the car as shown in Fig. 2. The height of the passenger cars is about 1.5m and about 1.9m for the SUVs. The RSEs installed at signalized intersections are placed about 8.5 m above the road surface and the receiver sensitivity is -94 dB.

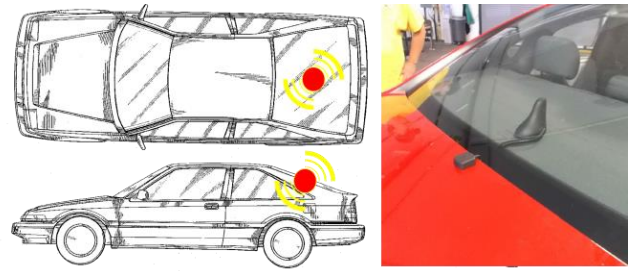

Fig. 2. ASD and VAD vehicle DSRC antenna installation location

During the experiment, BSM is transmitted by the vehicles and received and recorded by the RSE and DAS equipped vehicles. Sampled location coverage of BSM and probed vehicle data from one day is shown in Fig. 3 and Fig. 4. It can be seen that most vehicles remain in the city of Ann Arbor, and some of them are in the Southeastern Michigan or Northern

Ohio. By the end of June 2014, SPMD data covered 25 million miles of driving over 887 thousand hours [18].

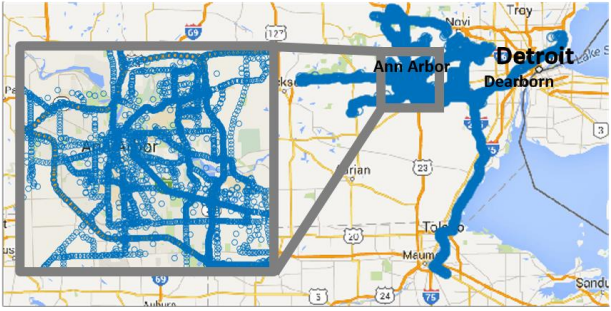
Fig. 3. BSM recorded by vehicles on May 1st, 2014

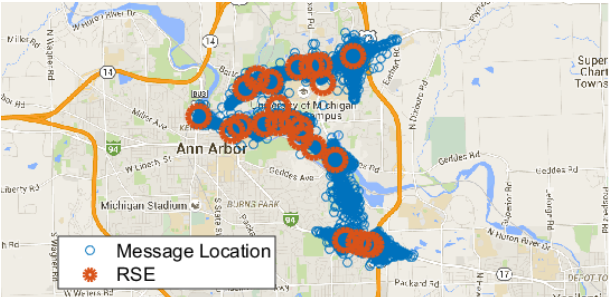
Fig. 4. Probed vehicle data recorded by RSEs on May 1st, 2014

B. Sampled Datasets

In this study we used vehicle data logged by the RSEs over 933 days from August 24th, 2012 to March 13th, 2015. The position, speed and heading information are used to study the performance of DSRC. 52 vehicles are equipped with ASD provided by Supplier A, 500 vehicles are equipped with VAD from Supplier B, and 500 vehicles are equipped with VAD from Supplier C. The following road sections are selected to study the influence factors of DSRC performance:

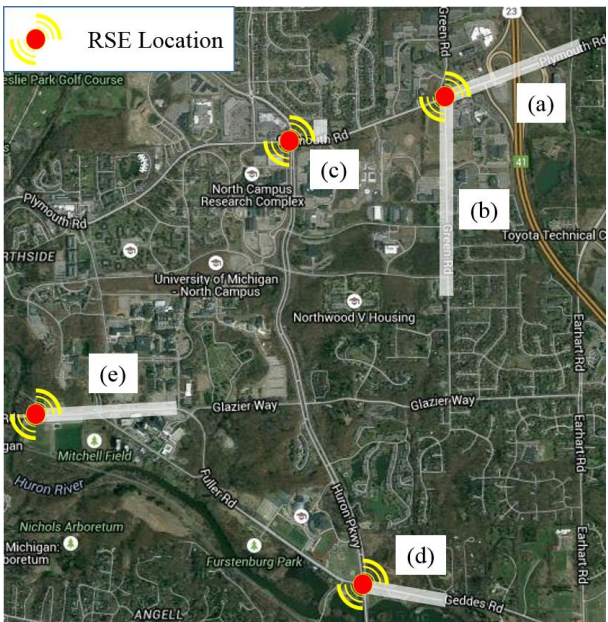
Fig. 5. Sampled locations with indicated routes

1) East and south legs of the Plymouth-Green intersection as shown in Fig. 5(a) and (b). Both of them are straight for about 2 km but have noticeable road elevation and foliage along the roadside.
2) Plymouth-Huron Parkway intersection as shown in Fig. 5(c), with non-line-of-sight (NLOS) blockage by a building northeast of the intersection.
3) East leg of the Huron Parkway-Geddes intersection as shown in Fig. 5 (d), with rows of trees at 400 m east of the RSE.
4) East leg of the Fuller-Cedar Bend Drive intersection as shown in Fig. 5 (e), which is straight for about 1 km with no foliage.

The locations of the studied RSEs are summarized in TABLE II.

TABLE II
Position Information of the Studied RSEs

| Location | Latitude | Longitude |
| --- | --- | --- |
| Fuller-Cedar Bend | 42.28714º | -83.7238º |
| Fuller-Geddes | 42.2776º | -83.699º |
| Plymouth-Huron Pkwy | 42.30258º | -83.7043º |
| Plymouth-Green | 42.30489º | -83.6926º |

Hourly weather record is obtained from the National Climatic Data Center [19] to study the influence of weather to DSRC performance. The weather data is recorded by a station located at Ann Arbor Municipal Airport. During the studied period, 22,622 hours of clear weather, 1,188 hours of rain, and 1,379 hours of snow are recorded. The amount of precipitation was not taken into consideration.

For each vehicle approaching the RSE of interest, data is recorded starting from the time the RSE receives the first message sent by the vehicle until the vehicle is within 20 meters to the RSE. For each vehicle leaving the RSE of interest, data is recorded starting from the time the vehicle is at 20 meters from the RSE, and ended at the last moment a message was heard. To calculate PDR, the distance to RSE is estimated using the vehicle speed and the number of packets not received by RSE is backfilled using time and estimated vehicle position. Then we divided the collected data into 10-meter distance bins. The data in the distance bins are used to estimate PDR as a function of distance from RSE.

The altitude of the subject road is measured using averaged GPS data. The data is available from test vehicles equipped with ASD and DAS. We assumed that the change of road altitude in the lateral direction is small compared with the change in longitudinal direction. Vehicles passing the subject road during the test period are all used. Then the road is divided into 1-meter bin. Only distance bins with more than 1000 data points are kept in the altitude calculation. The relative elevation between the road plane and the RSE is then computed.

IV. RESULTS

A. Performance under static obstruction

1) Influence of road elevation

Fig. 6 shows a sample measurement of the data accuracy using the data from the South leg of the Plymouth-Green intersection. The number of trips for this measurement is 36,665. The standard deviation of all road altitude data is 2.93m. We assume the average of each bin is an accurate

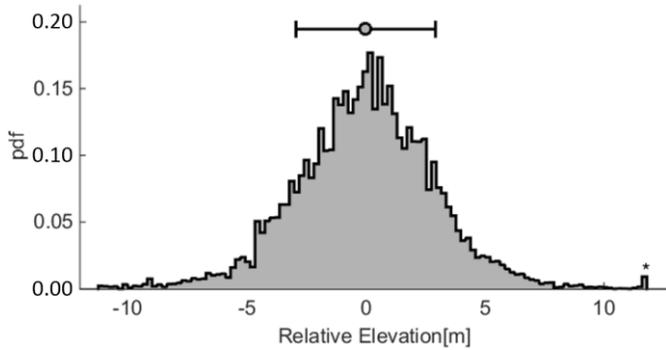

Fig. 6. Sample Relative Height Measurement from 400 ± 1 m to the south of Plymouth Road- Green Road Intersection

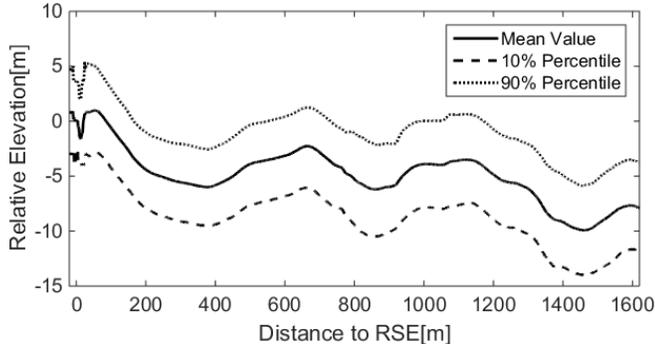

Fig. 7. Road Relative Elevation of the South leg of the Plymouth-Green intersection

measure of the road elevation. The computed results are shown in Fig. 7.

To evaluate the performance of DSRC with respect to road relative elevation only, we narrow down to data from the following subset: vehicles equipped with VAD from supplier B on both South leg and East leg of the Plymouth-Green intersection, clear weather, incoming, and at speed higher than 10m/s. The relative elevation profiles of these two road sections are shown in Fig. 8. We had minimized the effect of several other factors including the NLOS of buildings, tree foliage, weather, and DSRC hardware by careful selection of data.

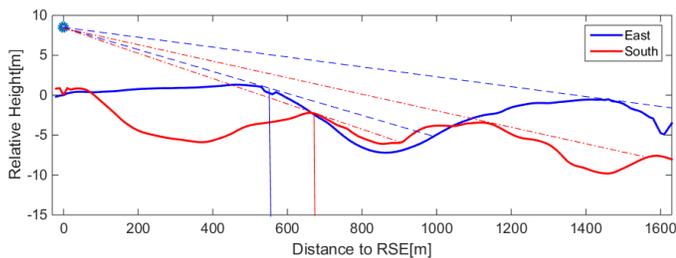

Fig. 8. Relative Elevation Profile of the East and South legs of the Plymouth-Green intersection

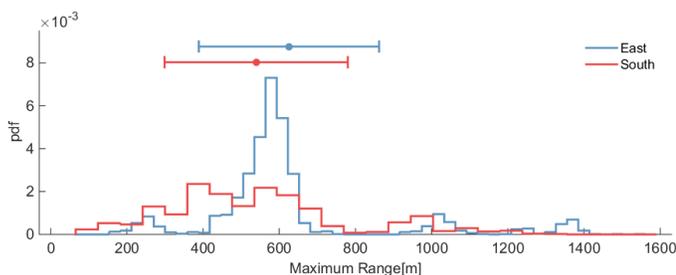

Fig. 9. Starting Distance of the East and South legs of the Plymouth-Green intersection

The dash lines in Fig. 8 show the line-of-sight path of the RSE. The RSE is installed at a height of about 8.5 m. The sample size is 4,117 for the South leg, and is 7,015 for the East leg. The probability density functions of the starting distance are shown in Fig. 9.

It can be seen that below 800 m, the road altitude has a noticeable effect on DSRC communication, as the highest peaks of the starting distance plots align with the line-of-sight and road relative elevation profile. For vehicles far from RSE, the weakening of signal due to NLOS can be a dominate factor for failing to communicate.

The PDRs are computed for distance up to 500m, for vehicle speed higher than 10m/s between November and February (light tree foliage), clear weather conditions. The PDRs of the two road sections are shown in Fig. 10.

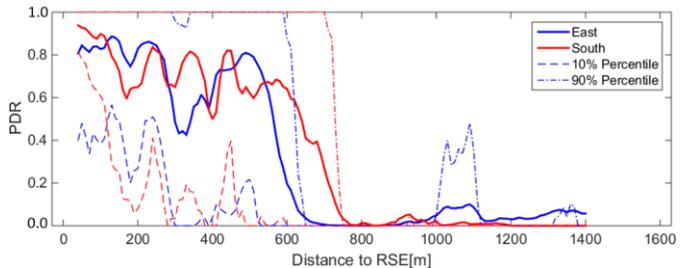

Fig. 10. PDR of the East and South legs of the Plymouth-Green intersection

The PDRs largely decrease with distance to an RSE, but have noticeable fluctuations. We do not yet have a complete explanation for the fluctuations, but believe there are at least three possible factors: First, the two-ray ground reflection model developed for flat roads in [20] demonstrates that a trough exists in PDR. Because roads are not flat in the real world and there are reflectors along the road such as other vehicles [21], multiple troughs exist. Secondly, the DSRC antenna is mounted inside the vehicle, thus the reflectors inside the vehicle would further contribute to the multipath effect [22]. And thirdly, other vehicles can enter/exit or stop at the (non-signaled) minor intersections or bus stop along the road, causing a higher concentration of vehicles at those locations and thus blockage and reflection which is a significant factor for PDR drop [7], [23].

2) Influence of NLOS Caused by Buildings

A major factor for DSRC communication is buildings NLOS blockage [20]. The Plymouth-Huron Pkwy intersection is selected as the focus site. The starting communication results and the line-of-sight blockage are shown in Fig. 11.

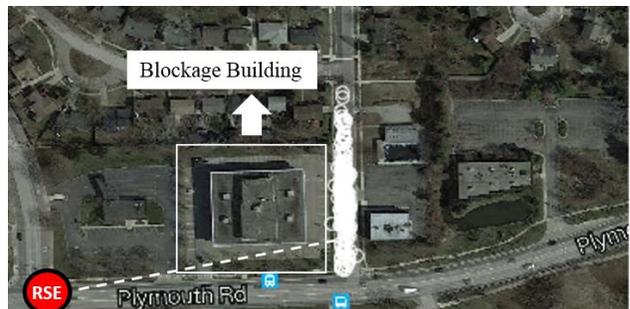

Fig. 11. Starting Points to the North of Plymouth-Huron Pkwy Intersection with Building Blockage

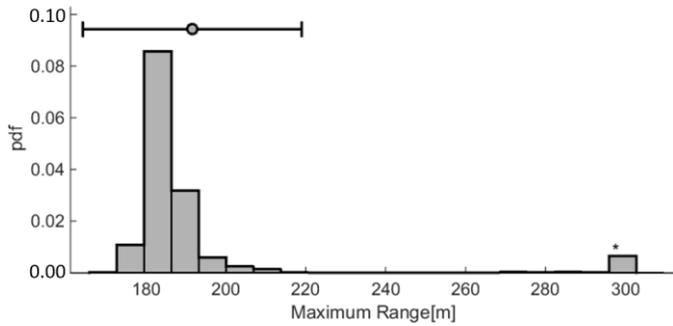

Fig. 12. Starting distance to the North of Plymouth-Huron Pkwy Intersection with Building Blockage

To highlight the effect of building blockage, we did not use data from vehicle traveling west on Plymouth. Instead, vehicles traveling south on the street about 170 meters east of the intersection are used. The starting points of communication on Fig. 11 shows building blockage would result in a blank space in communication starting points. Data from November to February is used to minimize the effect of tree foliage. Also, only clear weather conditions are considered. The sample size for this situation with DSRC from supplier B is 813. The starting distance under the NLOS condition is shown in Fig. 12. The starting distance varies from 170m to more than 300m.

However, only 30 out of 813 events yield a starting distance larger than 220m. This confirms that building blockage is a major factor in DSRC communication.

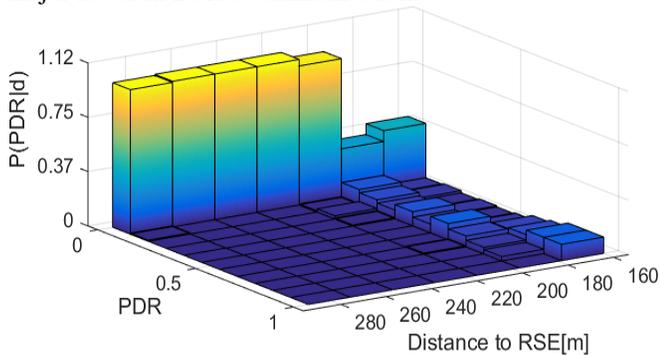

Fig. 13. Marginal Distribution of PDR to the North of Plymouth-Huron Pkwy Intersection with Building Blockage

Successful transmission under NLOS conditions is possible by diffraction and reflection from other buildings and vehicles. It can be seen from Fig. 13 that the NLOS PDR is much lower compared with the LOS cases in Fig. 10. This finding is consistent with the simulation results in [11], in which a knife edge diffraction model is used to simulate the behavior of NLOS transmission of DSRC.

3) Influence of NLOS Caused by Tree Foliage

Another common cause of NLOS conditions is tree foliage. The data used is from the east leg of the Huron Parkway-Geddes intersection, with foliage caused NLOS blockage starting from about 500m to RSE. We obtained "summer data" from May to August and "winter data" from November to February. Only clear weather, VAD from Supplier B are used in the analysis. The number of trips is 1,280 for winter, and 1,506 for summer. The starting points for the winter scenario are shown in Fig. 14.

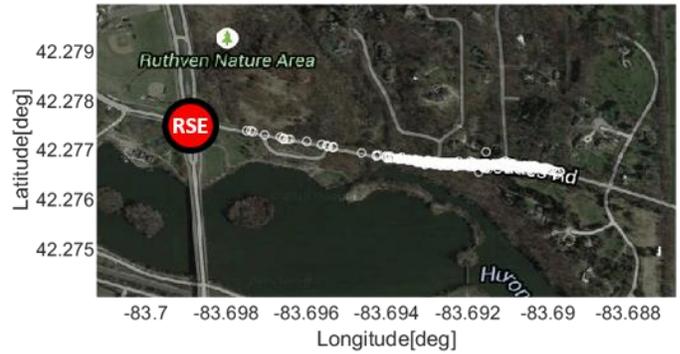

Fig. 14. Starting Points to the east of Fuller-Geddes Intersection with Foliage Blockage

The starting distance results are shown in Fig. 15. There is a small periodical behavior. During winter, the starting distance is slightly higher. In addition, there are many more outliers in the data from the summer months. The outliers of the whisker plots are data more than $\pm 2.7\sigma$ away from the mean value.

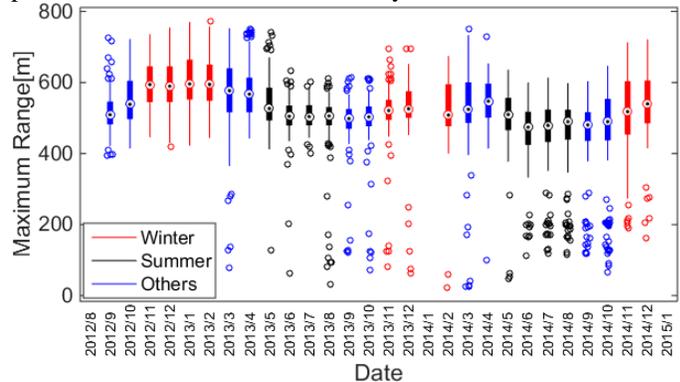

Fig. 15. Starting Distance results from the East leg of the Fuller-Geddes Intersection

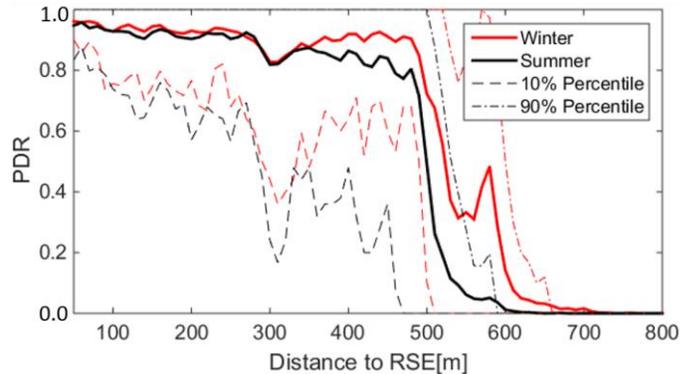

Fig. 16. PDRs to the east of Fuller-Geddes Intersection showing the effect of tree foliage

The PDRs of the winter and summer months are shown in Fig. 16. Unlike NLOS caused by buildings, the effect of tree foliage is modeled as attenuation-through-transmission [24], and the attenuation is the lumped result from reflections, diffractions, and scattering. It is also affected by the tree type (leaf shape, size and arrangement, deciduous or coniferous, etc.) and size. It is a significant task to build a comprehensive tree foliage attenuation model for DSRC. For this particular intersection, tree foliage reduces the effective range of DSRC by about 20 meters and reduces the PDR by up to 10 percentage points.

## B. Performance comparison of weather conditions

The effect of weather conditions is studied using data from the East leg of the Fuller-Cedar Bend intersection. The data used is from November to February, for vehicle speed higher than 5m/s. The sample size for clear weather is 2,581, 114 for rain, and 227 for snow. The starting points and starting distance are shown in Fig. 17.

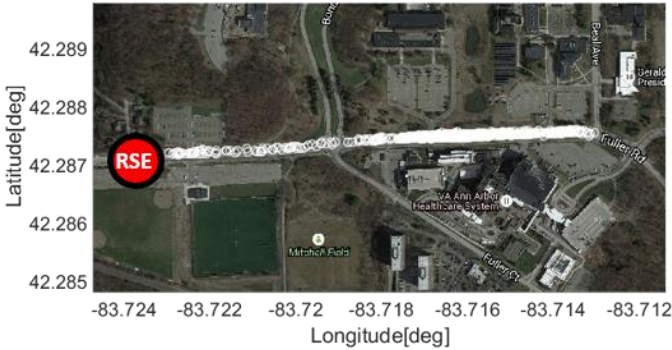

Fig. 17. Starting Points to the east of Fuller-Cedar Bend

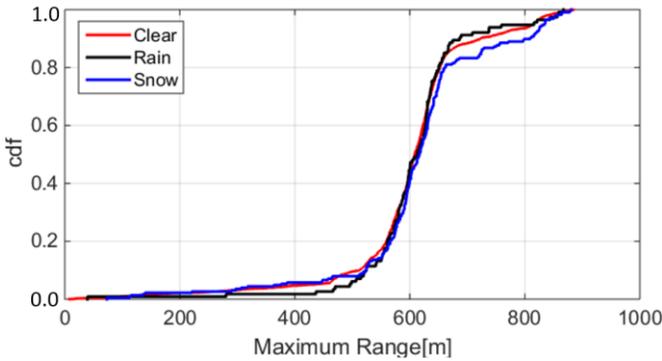

Fig. 18. Cumulative Distribution Function of starting distance to the East of Fuller-Cedar Bend

The cumulative distribution functions [25] of starting distance under different weather conditions are shown in Fig. 18. The maximum range is not noticeably affected by the weather condition.

To study the weather effect more rigorously, we apply the Kolmogorov–Smirnov (KS) test to the null hypothesis that the maximum range data for different weather conditions are drawn from the same population against its alternative hypothesis. Since the test is among three samples, and Kolmogorov–Smirnov test compares the cumulative distribution functions of two, Bonferroni correction is applied to correct the comparison among three samples. To reject the hypothesis at a significance level $\alpha$, for each individual hypothesis, the significance should be higher than

$$\alpha_i = \alpha / m \quad (1)$$

In the correction, $\alpha_i$ is the significance level for each hypothesis test, and $m$ is the number of hypotheses. Three hypotheses are tested in this situation.

The critical values is calculated from

$$c(\alpha) = c_0(\alpha) / \sqrt{\frac{n_1 n_2}{n_1 + n_2}} \quad (2)$$

TABLE III
Critical Values for Each Significance Level Used in KS test [26]

| $\alpha$ | 0.10 | 0.05 | 0.025 | 0.01 | 0.005 | 0.001 |
|---|---|---|---|---|---|---|
| $c_0(\alpha)$ | 1.22 | 1.36 | 1.48 | 1.63 | 1.73 | 1.95 |

Where $n_1$ and $n_2$ are sample sizes for the two tested samples. The test results for significance level 10% are summarized in TABLE IV, where D is the supremum of the difference between the empirical cumulative distribution functions of the two samples.

TABLE IV
KS test for starting distance under different weather conditions

| Hypothesis | D | c(α) |
|---|---|---|
| Clear vs. Rain | 0.0545 | 0.137 |
| Clear vs. Snow | 0.0747 | 0.099 |
| Snow vs. Rain | 0.0980 | 0.164 |

Based on the results from the KS test, with all testing statistic values (D) less than the threshold values ($c(\alpha)$), the null hypothesis cannot be rejected. In other words, the starting distance behavior under different weather conditions are not distinguishable.

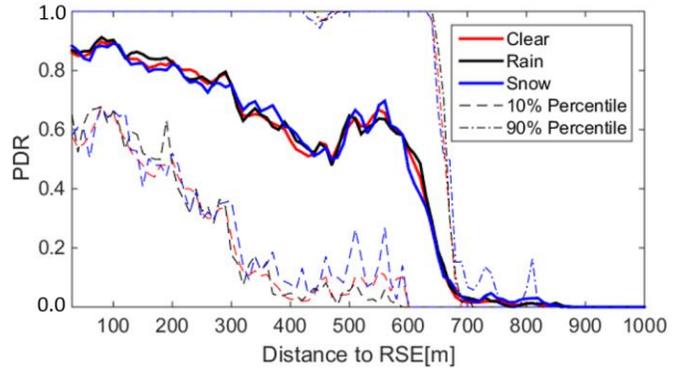

Fig. 19. Overall PDR from the east leg of Fuller-Cedar Bend intersection

The PDRs under different weather conditions are shown in Fig. 19. It can be seen that the PDRs under the three weather conditions are close to each other. We conclude that weather conditions have little influence on the performance of DSRC in terms of both of maximum range and PDR.

## C. Performance comparison for vehicle moving direction

The effect of vehicle moving direction on DSRC performance is studied using data from the South leg of the Plymouth-Green intersection. Data from November to February under clear weather conditions with vehicle speed at the starting/breaking points higher than 15m/s are used. The sample size is 1,179 outgoing, and 877 for incoming. The starting points are shown in Fig. 20.

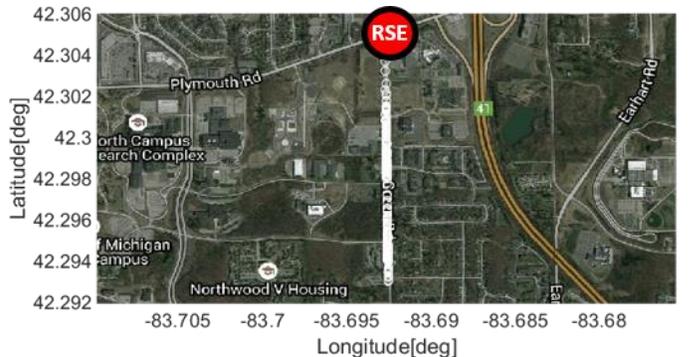

Fig. 20. Starting Points to the south of Fuller-Green intersection

The small peak in breaking distance at about 700m in the outgoing cases is due to road elevation as discussed earlier and the peak at about 1200m is also affected by road blockage. It is clear that the maximum range of the outgoing cases is much higher than that for the incoming cases. This finding is consistent with the experimental results from [27].

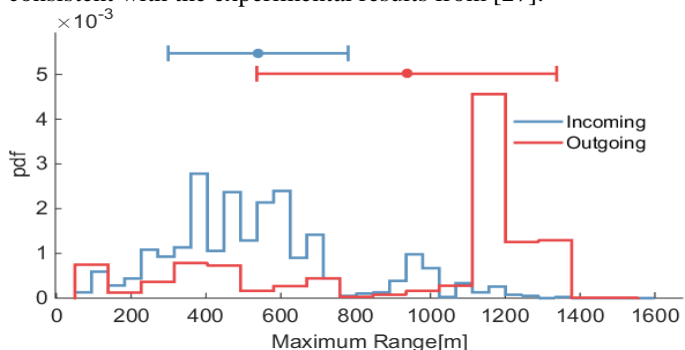
Fig. 21. Maximum Range to the South of Plymouth-Green intersection

The PDR results are shown in Fig. 22. The PDR of the incoming cases is consistently lower than that of the outgoing cases. Since controlled experiment by Bai et al [6] found that mobility shouldn't cause performance difference, we believe this is due to the location of the DSRC antenna: under the rear window for our vehicles. This is consistent with the findings from Mincic et al. [22]. In their study the antenna is placed on the roof top, dashboard and below the rear mirror. The PDR to different directions of the vehicle at a transmission distance of 61 meters are compared. Their study shows that the antenna on the roof top has the best performance with PDR above 95% in all directions. The antenna on the dashboard has a PDR around 70% to the front of vehicle and below 20% to the back. This indicates that the antenna should be placed at the front of the vehicles for better communication with vehicles/RSEs in front of the vehicle. Placing the antenna at the back could reduce PDR on average by 10-20%.

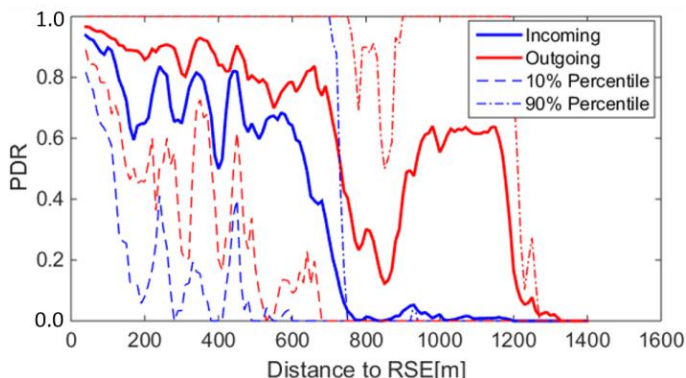
Fig. 22. PDR on the South leg of the Plymouth-Green intersection

D. Effect of vehicular blockage

The effect of attenuation due to other vehicles is studied using data from the South leg of Plymouth-Green intersection with DSRC from Supplier B during winter time. The sample size for every hour of the day is shown in Fig. 23. The morning and evening rush hours are around 8:00 and 17:00 respectively.

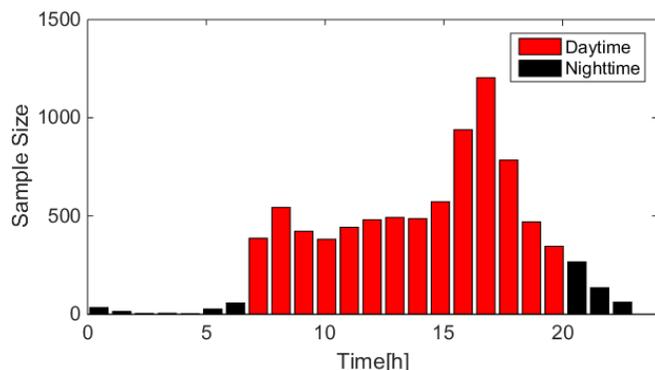
Fig. 23. Sample Size of Maximum Range for Different Time on the South leg of the Plymouth-Green intersection

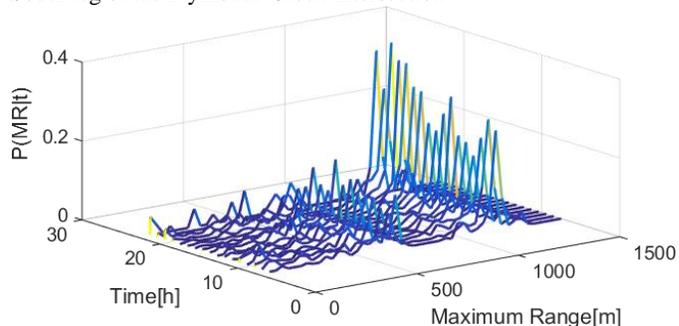
Fig. 24. Marginal Probability Distribution of Maximum Range with respect to time on the South leg of the Plymouth-Green intersection

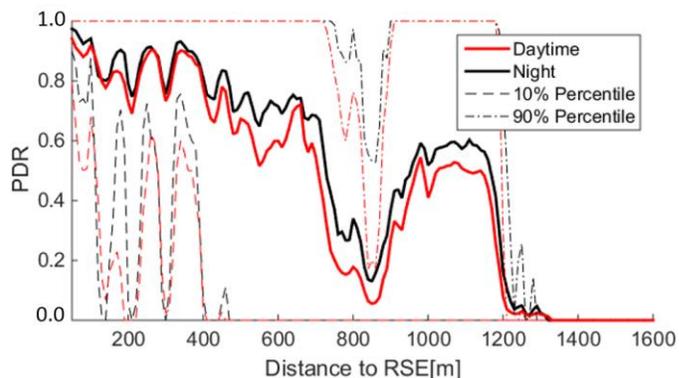
Fig. 25. Overall PDR of Daytime and Night on the South leg of the Plymouth-Green intersection

This road section is relatively flat as shown in Fig. 8. The maximum range at the different time of the day is shown in Fig. 24.

As shown in Fig. 24, during late night, from 10 pm to 2 am, the mean value of the maximum range is higher than that during the daytime. This indicates that the DSRC maximum range is affected by surrounding traffic. A controlled test on a parking lot was done in [7], which concludes that with a truck in between two cars communicating through DSRC, the transmission power level can reduce by about one-third compared to the no-blockage case. If daytime is defined as 7:00 to 20:00, and nighttime as 21:00 to 6:00, with 482 samples for daytime and 80 samples for the nights. The PDR is shown in Fig. *25*. PDR during the daytime is lower than that in the night, likely due to the NLOS caused by the presence of other vehicles

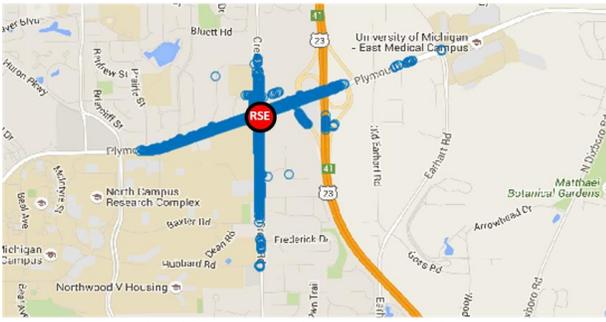

Fig. 26. BSM received by RSE during 8:27:14-8:37:12, Jul. 29th, 2013

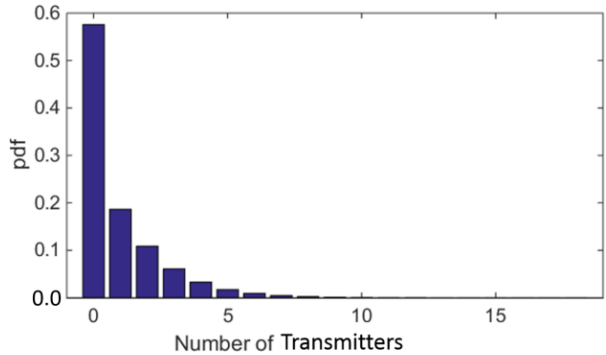

Fig. 27. Number of transmitters during the 3-year experiment

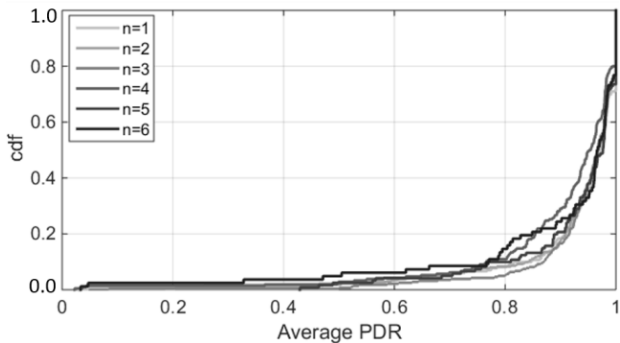

Fig. 28. Average PDR within 100m with different number of transmitters on the South leg of the Plymouth-Green intersection

E.  Effect of transmitter number

The performance of DSRC with different number of transmitters is studied with data from all SUVs equipped with VAD from supplier B moving on the south leg of Plymouth-Green Intersection. The number of transmitters is counted as the total number of vehicles communicating with RSE using a 1-second time bin. Based on the assumption that packet collision would dominate the cause of packet loss within 100m, the average of PDR within 100m of each sample is used to study the effect of transmitter number.  Fig. 26 shows a 10-minute sample of BSM received by RSE, each point in the plot indicates a received BSM. Fig. 27 shows the probability density function of transmitter number during the 3-year experiment. Safety Pilot Model Deployment consists more than 2800 vehicles, which is about 2.5 % of the total traffic in Ann Arbor, thus more than half of the time the number of transmitters is 0 and the maximum number of transmitting nodes at the same time is 17 for the studied RSE.

The cumulative distribution function of average PDR is shown in Fig. 28. Only samples with a size larger than 50 were considered in the analysis. KS-test is applied to the null hypothesis that the average PDR distribution with transmitter number 1 and 6 are drawn from the same population against its alternative hypothesis at a significance level 10%. The supremum of the difference between the empirical cumulative distribution functions of the two samples is 0.1092, and the threshold at 10% significance level is 0.1713 after corrected with the sizes of the two samples. This indicates that the average PDR of transmitter number 1 and 6 cannot be distinguished statistically. However, the cumulative distribution function plot shows a slightly increase in variance of average PDR with the increase of transmitter number.

Our result is consistent with the experiment findings from Ramachandran et al. [8], which found the largest performance drop due to packet collision happens with more than 30 transmitters present. Considering the low penetration ratio of DSRC equipped vehicle in Safety Pilot Model Deployment, the packet collision is not a major cause of packet loss.

F.  Performance of DSRC from different suppliers

The performance of DSRC systems from different suppliers is compared using data from all sedan on a straight road with few trees (South leg of Plymouth-Green) during May to August, under clear weather conditions, and with vehicle speed higher than 10 m/s. The transmitting power of DSRC from the suppliers are summarized in TABLE V. The values are close to each other and less than 33 dBm which is the maximum transmitting power of 5.8 GHz signal defined by IEEE 802.11p[3]. The maximum range results are shown in Fig. 29.

TABLE V
DSRC Transmitting Power

| Supplier | Maximum Output Power(dBm) |
|---|---|
| A | 22 |
| B | 21 |
| C | 22 |

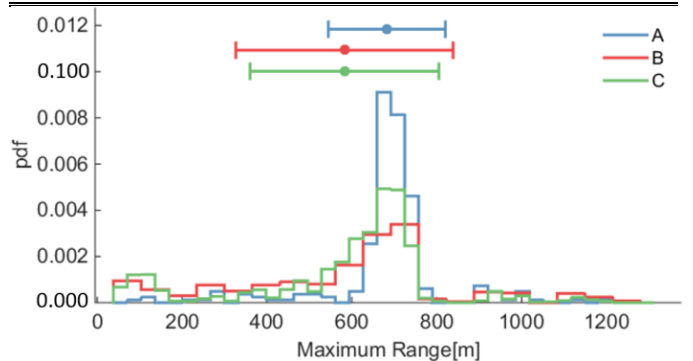

Fig. 29. Maximum Range of DSRC from three suppliers using data from the south leg of Plymouth-Green

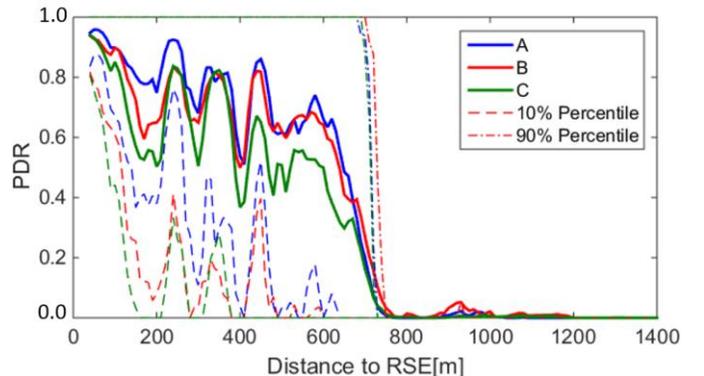

Fig. 30. PDR of DSRC from three suppliers using data from the south leg of Plymouth-Green

In terms of maximum range, DSRC radios from all three suppliers show some variation but the average values are similar—around 600 meters. The PDR from data between November and April (Fig. 30) show some difference. Most of the time the PDR are above 70% for range below 150 meters. However the 10% percentile results are pretty low, showing that sometimes the DSRC communication is not effective. Possible factors include hardware or software malfunctions, influence from a large vehicle (bus, truck), antenna location and installation, snow accumulation covering the RSE antenna and its effect on ground reflection.

## V. CONCLUSION

In this paper we presented real-world performance of DSRC using a large set of naturalistic driving data obtained through the University of Michigan Safety Pilot Model Deployment project. We mainly focus on the maximum range and packet delivery ratio of V2I communication between 1,050 vehicles and selected road-side equipment (RSE).

Our analysis results show that the most influential factors to the maximum range and PDR include: non-line-of-sight (NLOS) obstruction from static (e.g., buildings) and moving objects (e.g., vehicles). The location of the antenna in the vehicle also affects the range and PDR noticeably. Different weather conditions show little influence on the performance of DSRC and the number of transmitter shows little influence with transmitter number less than 6.

Our analysis results largely agree with earlier studies from the literature. Controlled lab experiments are better suited to study particular influential factors, whereas the naturalistic driving data from a large fleet such as what we used show the lumped effect of multiple factors in real-world driving. Cross examination of the results from both types of studies can guide the design, installation and deployment of DSRC so that they can operate more reliably in the future.

## VI. ACKNOWLEDGEMENT

The authors would like to acknowledge the help from Scott Bogard on database query, and Dr. James R. Sayer and Ms. Debra Bezzina on the general support for access of the Safety Pilot Model Deployment data. The authors would also like to thank the anonymous reviewers for their valuable comments and suggestions to improve the quality of the paper.

## REFERENCES

[1] Dedicated Short Range Communications (DSRC) Message Set Dictionary, SAE J2735, Apr. 2015
[2] "Intelligent Transport Systems (ITS); Vehicular Communications; Basic Set of Applications; Part 2: Specification of Cooperative Awareness Basic Service," Tech. Rep. EN 302 637-2, 2014.
[3] Draft Amendment for Wireless Access in Vehicular Environments (WAVE), IEEE P802.11p/D3.0, July 2007.
[4] H. Hartenstein. "Physical Layer Considerations for Vehicular Communications," VANET: vehicular applications and inter-networking technologies. Vol. 1. United Kingdom.
[5] M. Boban and P.M. d'Orey, "Measurement-based evaluation of cooperative awareness for V2V and V2I communication," Vehicular Networking Conference (VNC), 2014 IEEE, vol., no., pp.1,8, 3-5 Dec. 2014
[6] F. Bai, D. D. Stancil and H. Krishnan "Toward understanding characteristics of dedicated short range communications (DSRC) from a perspective of vehicular network engineers", Proc. 16th Annu. Int. Conf. MobiCom, pp.329 -340 2010
[7] R. Meireles , M. Boban , P. Steenkiste , O. Tonguz and J. Barros "Experimental study on the impact of vehicular obstructions in VANETs", Proc. 2nd IEEE VNC, pp.338 -345 2010
[8] K. Ramachandran, M. Gruteser, R. Onishi, T. Hikita, "Experimental analysis of broadcast reliability in dense vehicular networks," Vehicular Technology Magazine, IEEE , vol.2, no.4, pp.26,32, Dec. 2007
[9] H. Schumacher and H. Tchouankem, "Highway Propagation Modeling in VANETS and Its Impact on Performance Evaluation," in IEEE/IFIP WONS 2013. Banff, Canada: IEEE, Mar. 2013, pp. 178-185.
[10] X. Ma, J. Zhang and T. Wu "Reliability analysis of one-hop safety-critical broadcast services in VANETs", IEEE Trans. Veh. Technol., vol. 60, no. 8, pp.3933 -3946 2011
[11] S. Biddlestone, K. Redmill, R. Miucic and Ü. Özgüner "An integrated 802.11 p WAVE DSRC and vehicle traffic simulator with experimentally validated urban (LOS and NLOS) propagation models", IEEE Trans. Intell. Transp. Syst., vol. 13, no. 4, pp.1792 -1802 2012
[12] J. Fernandez , K. Borries , L. Cheng , V. Bhagavatula , D. Stancil and F. Bai "Performance of the 802.11p physical layer in vehicle-to-vehicle environments", IEEE Trans. Veh. Technol., vol. 61, no. 1, pp.3 -14 2012
[13] M. Hassan, H. Vu and T. Sakurai, "Performance Analysis of the IEEE 802.11 MAC Protocol for DSRC Safety Applications", IEEE Trans. Vehicular Technology, vol. 60, no. 8, pp. 3882-3896, Oct. 2011.
[14] K. A. Hafeez , L. Zhao , B. Ma and J. W. Mark "Performance analysis and enhancement of the DSRC for VANET's safety applications", IEEE Tran. Veh. Technol., vol. 62, no. 7, pp.3069 -3083 2013
[15] J. S. Hickman and R. J. Hanowski "An assessment of commercial motor vehicle driver distraction using naturalistic driving data", Traffic Inj. Prev., vol. 13, no. 6, pp.612 -619 2012
[16] J.S. Hickman, F. Guo, M.C. Camden, R. J. Hanowski, A. Medina and J. E. Mabry, "Efficacy of roll stability control and lane departure warning systems using carrier-collected data", J. Safety Research, vol. 52, pp. 59–63, Feb. 2015.
[17] M. Dozza, J. Bargman and J.D. Lee, "Chunking: A procedure to improve naturalistic data analysis," Accident Analysis & Prevention, vol. 58, pp. 309–317, Sept. 2013.
[18] D. Bezzina and J. Sayer, "Safety pilot model deployment: Test conductor team report," Report No. DOT HS, vol. 812, p. 171.
[19] National Climatic Data Center (NCDC) Integrated Surface Global Hourly Data, Available at https:// data.noaa.gov / dataset / integrated - surface – global - hourly - data
[20] M. Boban, J. Barros and O.K. Tonguz, "Geometry - Based Vehicle- to-Vehicle Channel Modeling for Large-Scale Simulation," Vehicular Technology, IEEE Transactions on , vol.63, no.9, pp.4146,4164, Nov. 2014
[21] J. Harri, H. Tchouankem, O. Klemp, and O. Demchenko, "Impact of vehicular integration effects on the performance of DSRC communications," in *Wireless Communications and Networking Conference (WCNC), 2013 IEEE*, 2013, pp. 1645-1650.
[22] R. Miucic and S. Bai, "Performance of Aftermarket (DSRC) Antennas Inside a Passenger Vehicle", SAE Int., vol. 4, no. 1, pp. 150–155, Apr. 2011.
[23] J. Gozálvez, M. Sepulcre, and R. Bauza, "IEEE 802.11 p vehicle to infrastructure communications in urban environments," Communications Magazine, IEEE, vol. 50, pp. 176-183, 2012.
[24] J. Goldhirsh and W. J. Vogel, "Handbook of propagation effects for vehicular and personal mobile satellite systems—Overview of experimental and modeling results", Appl. Phys. Lab., Johns Hopkins Univ./Elect. Eng. Res. Lab., Univ. Texas Austin, Austin, TX, USA, Tech. Rep. A2A-98-U-0-021 (APL), EERL-98-12A (EERL), Dec. 1998.
[25] W. C. Navidi, Statistics for engineers and scientists vol. 1: McGraw-Hill New York, 2006.
[26] M.H. Gail and S.B. Green, "Critical Values for the One-Sided Two-Sample Kolmogorov-Smirnov Statistic", J. American Statistical Association, vol. 71, no. 355, pp. 757–760, June 1975.

[27] S. Andrews and M. Cops, "Final Report: Vehicle Infrastructure Integration Proof of Concept Results and Findings Summary – Vehicle", USDOT, Novi, MI, Tech. Rep. FHWA-JPO-09-043, May 2009.

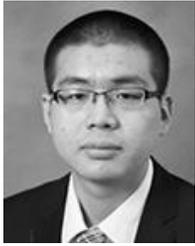

**Xianan Huang** received B.S. degree in mechanical engineering from Shanghai Jiaotong University and Purdue University in 2014. He is currently pursuing the Ph.D degree in mechanical engineering at University of Michigan, Ann Arbor.

From 2013 to 2014 he was an undergraduate researcher at Purdue University. Since 2014 he has been a graduate researcher at University of Michigan, Ann Arbor. His research interests include connected vehicle, hybrid vehicle, and system dynamics and controls.

Mr. Huang's awards and honors include A-Class scholar of Shanghai Jiaotong University and Summer Undergraduate Research Fellowship (Purdue University)

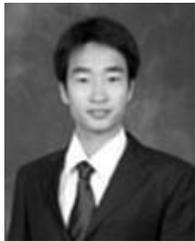

**Ding Zhao** received B.S. degree in automotive engineering from Jilin University, China in 2010. He is currently pursuing the Ph.D. degree in mechanical engineering at University of Michigan, Ann Arbor.

From 2011 to 2014 he was a Graduate Researcher in Ford - UM alliance project. Since 2012, he has been a Graduate Researcher UMTRI. His research interest includes automated vehicles, connected vehicles, driver modeling, crash avoidance, electric/hybrid vehicles, heavy trucks, and big data analysis.

Mr. Zhao's awards and honors include National Scholarships (Ministry of Education of the P.R. of China), Red Flag scholarship (First Automobile Works Group Corporation), Summa Cum Laude (Jilin University), and Departmental Fellowship (University of Michigan).

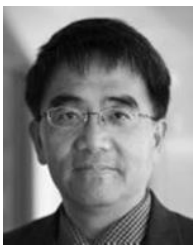

**Huei Peng** received the Ph.D. degree from the University of California, Berkeley, CA, USA, in 1992. He is currently a Professor with the Department of Mechanical Engineering, University of Michigan, Ann Arbor. He is currently the U.S. Director of the Clean Energy Research Center—Clean Vehicle Consortium, which supports 29 research projects related to the development and analysis of clean vehicles in the U.S. and in China. He also leads an education project funded by the Department of Energy to develop ten undergraduate and graduate courses, including three laboratory courses focusing on transportation electrification. He serves as the Associate Director of the University of Michigan Mobility Transformation Center, a center that studies connected and autonomous vehicle technologies and promotes their deployment. He has more than 200 technical publications, including 85 in refereed journals and transactions. His research interests include adaptive control and optimal control, with emphasis on their applications to vehicular and transportation systems. His current research focuses include design and control of electrified vehicles, and connected/automated vehicles.